\documentclass[11pt,twoside]{article}
\usepackage{asp2014}

\aspSuppressVolSlug
\resetcounters
\begin{document}

\title{Preliminary Results of a Deep Learning Anomaly Detection Method to Identify Gamma-Ray Bursts in the AGILE Anticoincidence System
}

\author{N.~Parmiggiani,$^1$ A.~Bulgarelli,$^1$ A.~Ursi,$^2$, M.~Tavani,$^3$ A.~Macaluso,$^4$ A.~Di~Piano,$^1$ V.~Fioretti,$^1$ L.~Baroncelli,$^1$ A.~Addis,$^1$ and C.~Pittori$^{3,5}$}
\affil{$^1$INAF OAS Bologna, Via P. Gobetti 93/3, 40129 Bologna, Italy. \email{nicolo.parmiggiani@inaf.it}}

\affil{$^2$INAF/IAPS Roma, Via del Fosso del Cavaliere 100, I-00133 Roma, Italy.}

\affil{$^3$INAF/OAR Roma, Via Frascati 33, I-00078 Monte Porzio Catone, Roma, Italy.}

\affil{$^4$German Research Center for Artificial Intelligence (DFKI), 66123 Saarbruecken, Germany.}

\affil{$^5$ASI/SSDC Roma, Via del Politecnico snc, I-00133 Roma, Italy.}


\paperauthor{Nicol\`{o}~Parmiggiani}{nicolo.parmiggiani@inaf.it}{0000-0002-4535-5329}{INAF}{OAS}{Bologna}{BO}{40129}{Italy}
\paperauthor{Andrea~Bulgarelli}{andrea.bulgarelli@inaf.it}{0000-0001-6347-0649}{INAF}{OAS}{Bologna}{BO}{40129}{Italy}
\paperauthor{Alessandro~Ursi}{alessandro.ursi@inaf.it}{0000-0002-7253-9721}{INAF}{IAPS}{Roma}{RO}{00133}{Italy}
\paperauthor{Marco~Tavani}{marco.tavani@inaf.it}{0000-0003-2893-1459}{INAF}{OAR}{Monte Porzio Catone}{RO}{00078}{Italy}
\paperauthor{Antonio~Macaluso}{antonio.macaluso.90@gmail.com }{0000-0002-1348-250X}{}{DFKI}{Saarbruecken}{66123}{}{Germany}
\paperauthor{Ambra~Di~Piano}{ambra.dipiano@inaf.it}{0000-0002-9894-7491}{INAF}{OAS}{Bologna}{BO}{40129}{Italy}
\paperauthor{Valentina~Fioretti}{valentina.fioretti@inaf.it}{0000-0002-6082-5384}{INAF}{OAS}{Bologna}{BO}{40129}{Italy}
\paperauthor{Leonardo~Baroncelli}{leonardo.baroncelli@inaf.it}{0000-0002-9215-4992}{INAF}{OAS}{Bologna}{BO}{40129}{Italy}
\paperauthor{Antonio~Addis}{antonio.addis@inaf.it}{0000-0002-0886-8045}{INAF}{OAS}{Bologna}{BO}{40129}{Italy}
\paperauthor{Carlotta~Pittori}{carlotta.pittori@inaf.it}{0000-0001-6661-9779}{INAF}{OAR}{Monte Porzio Catone}{RO}{00078}{Italy}





\begin{abstract}

AGILE is a space mission launched in 2007 to study X-ray and gamma-ray astronomy. The AGILE team developed real-time analysis pipelines to detect transient phenomena such as Gamma-Ray Bursts (GRBs) and to react to external science alerts received by other facilities. The AGILE anti-coincidence system (ACS) comprises five panels (four lateral and one on the top) that surround the AGILE detectors to reject background charged particles. It can also detect hard X-ray photons in the energy range 50 - 200 KeV. The acquisition of the ACS data produces a time series for each panel. These time series can be merged in a single multivariate time series (MTS). We present in this work a new Deep Learning model for GRBs detection in the MTSs, generated by the ACS, using an anomaly detection technique. The model is implemented with a Deep Convolutional Neural Network autoencoder architecture. We trained the model with an unsupervised learning algorithm using a dataset of MTSs randomly extracted from the AGILE ACS data. The reconstruction error of the autoencoder is used as the anomaly score to classify the MTS. If the anomaly score is higher than a predefined threshold, the MTS is flagged as a GRB. The trained model is evaluated using a list of MTSs containing GRBs. The tests confirmed the model's ability to detect transient events, providing a new promising technique to identify GRBs in the ACS data that can be implemented in the AGILE real-time analysis pipeline. 
  
\end{abstract}

\section{Introduction}

AGILE (Astrorivelatore Gamma ad Immagini LEggero - Light Imager for Gamma-Ray Astrophysics) is a space mission of the Italian Space Agency (ASI) devoted to high-energy astrophysics and launched on 23rd Apr 2007 \citep{2008NIMPA.588...52T, 2009A&A...502..995T}. The AGILE payload consists of the Silicon Tracker (ST), the SuperAGILE X-ray detector, the CsI(Tl) Mini-Calorimeter (MCAL), and an AntiCoincidence System (ACS). The combination of ST, MCAL, and ACS composes the Gamma-Ray Imaging Detector (GRID). The ACS are composed of five independent panels (four lateral and one on the top) surrounding the AGILE detectors. The primary role of the ACS is to reject charged background particles. It can also detect hard X-ray photons in the energy range 50 - 200 KeV. The number of events/s that hit the ACS panels are continuously recorded in telemetry as ratemeters (RMs) data, with 1 second resolution, to monitor the high-energy background through the orbital phase. Each ACS panel RM count rate constitutes a time series. The time series of all five ACS panels can be considered a single multivariate time series (MTS) due to time-alignment. An MTS has more than one time-dependent variable. In this context, we expect that a transient event is detected by more than one panel simultaneously, increasing its detectability. 

The AGILE team developed real-time analysis (RTA) pipelines \citep{2019ExA....48..199B, Parmiggiani:20214o} to detect transient phenomena such as Gamma-Ray Bursts (GRBs) and to react to external science alerts sent by other facilities.\ This automated software system implements different algorithms to analyse data. This work aims to develop a new detection method based on Deep Learning (DL) to identify GRBs inside ACS data, which will be implemented into the AGILE RTA pipelines. Although the ACS was not intended for GRBs detection, with this work we aim to use this new DL method to improve the overall AGILE capability of identifying transient phenomena. The AGILE Team has already used DL techniques to detect GRBs, such as the detection pipeline developed for the GRID detector \citep{2021ApJ...914...67P}.

\section{Deep Learning Model}

We developed a DL model based an anomaly detection technique to identify MTSs significantly deviating from the background-only MTSs used for the training. Among the DL architectures able to execute this type of analysis \citep{10.1145/3439950}, we decided to implement our model with a Convolutional Neural Network (CNN) autoencoder \citep{Goodfellow-et-al-2016} using several 1D CNN layers that are designed to work with time series having one or more channels. The CNN are well known in several fields for the successes obtained with image processing (e.g. object detection and object segmentation). 

The autoencoders are neural networks designed to encode the input data in a representation with reduced dimension and then decode the compressed information to the original data, minimising the reconstruction error (the differences between the original input data and reconstructed one). The autoencoder can be trained with unsupervised techniques avoiding manual feature engineering for labelling the data. This kind of neural network can be used for anomaly detection because when a trained autoencoder receives as input an object different from whom present in the training dataset, it cannot efficiently reconstruct such object, resulting in a large reconstruction error. Our model comprises two 1D CNN layers for the encoding (with 250 and 200 filters, respectively) and two 1D CNN layers for the decoding (with 200 and 250 filters, respectively). Each 1D CNN layer uses the ReLu activation function followed by a Dropout layer with a 20$\%$ value. The network is implemented using two open-source frameworks, Keras  (\url{https://keras.io}) running on top of Tensorflow (\url{https://www.tensorflow.org}).

We prepared a dataset extracting 5000 MTSs, with a length of 40 seconds and bins of one second, from the ACS data archive. From this dataset, we excluded the time windows where the AGILE detectors were in idle mode, due to ongoing passages into the South Atlantic Anomaly (SAA), or where there are known GRBs. We apply these filters to ensure that the dataset contains background-only MTSs. To improve the training procedures, the MTSs are normalised. From this dataset, we obtain three datasets for the training, test, and validation phases. 

The training is performed with a batch size of 100 for 53 epochs, after which the reconstruction error plateaus and the procedure automatically stops to avoid overfitting. The optimisation algorithm is Adam \citep{2014arXiv1412.6980K} configured with a learning rate of 0.001, and the reconstruction error is computed as the mean squared error. 

We use the anomaly score (reconstruction error) to classify MTSs and detect anomalies. If the anomaly score is higher than a predefined threshold, that time window is flagged as an anomaly (e.g. a GRB detection). The threshold value is defined as the 99.8 percentile of the reconstruction error distribution obtained evaluating all MTSs in the test dataset with the model.

\begin{figure*}[!htb]
	\centering
	  \includegraphics[width=0.7\textwidth]{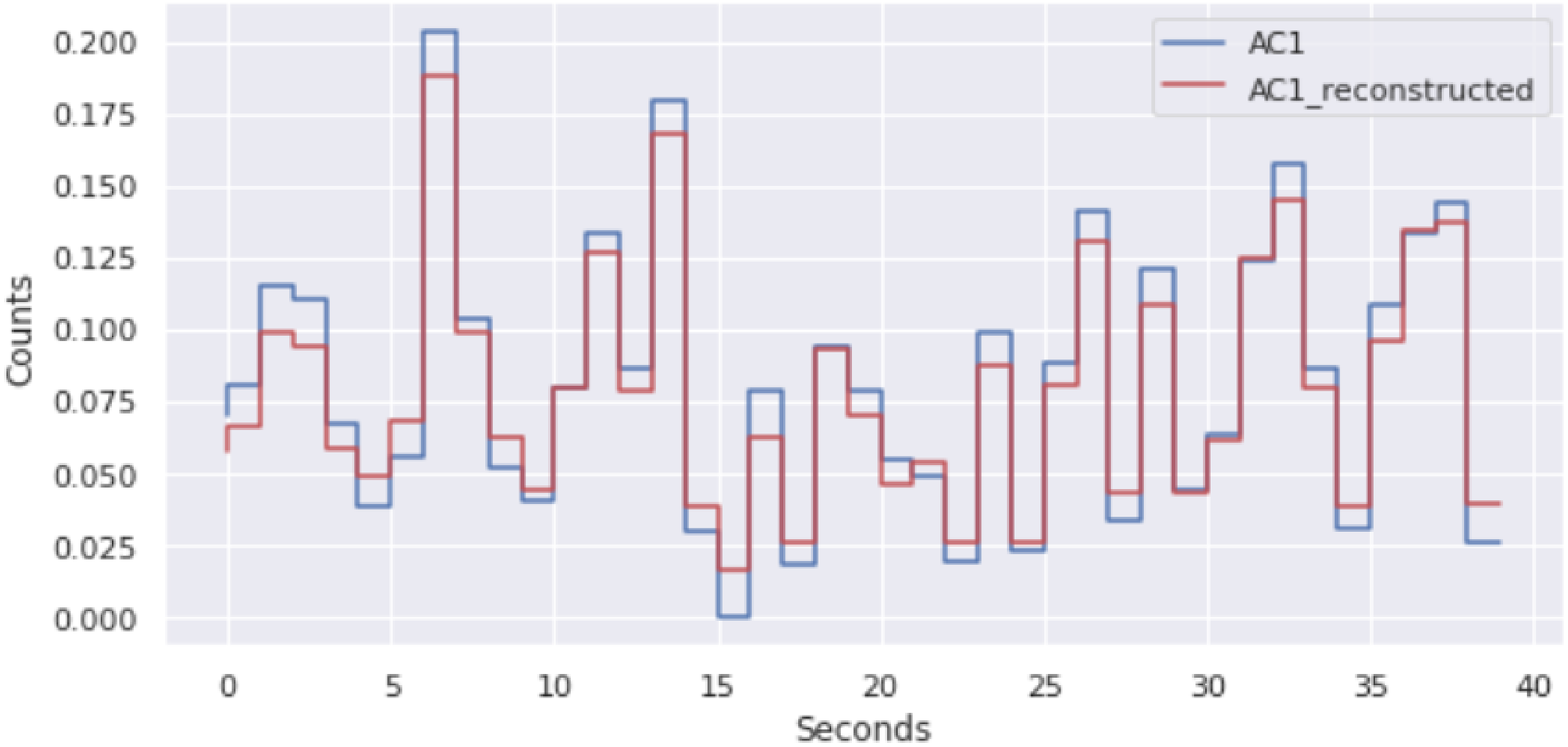}
	  \includegraphics[width=0.7\textwidth]{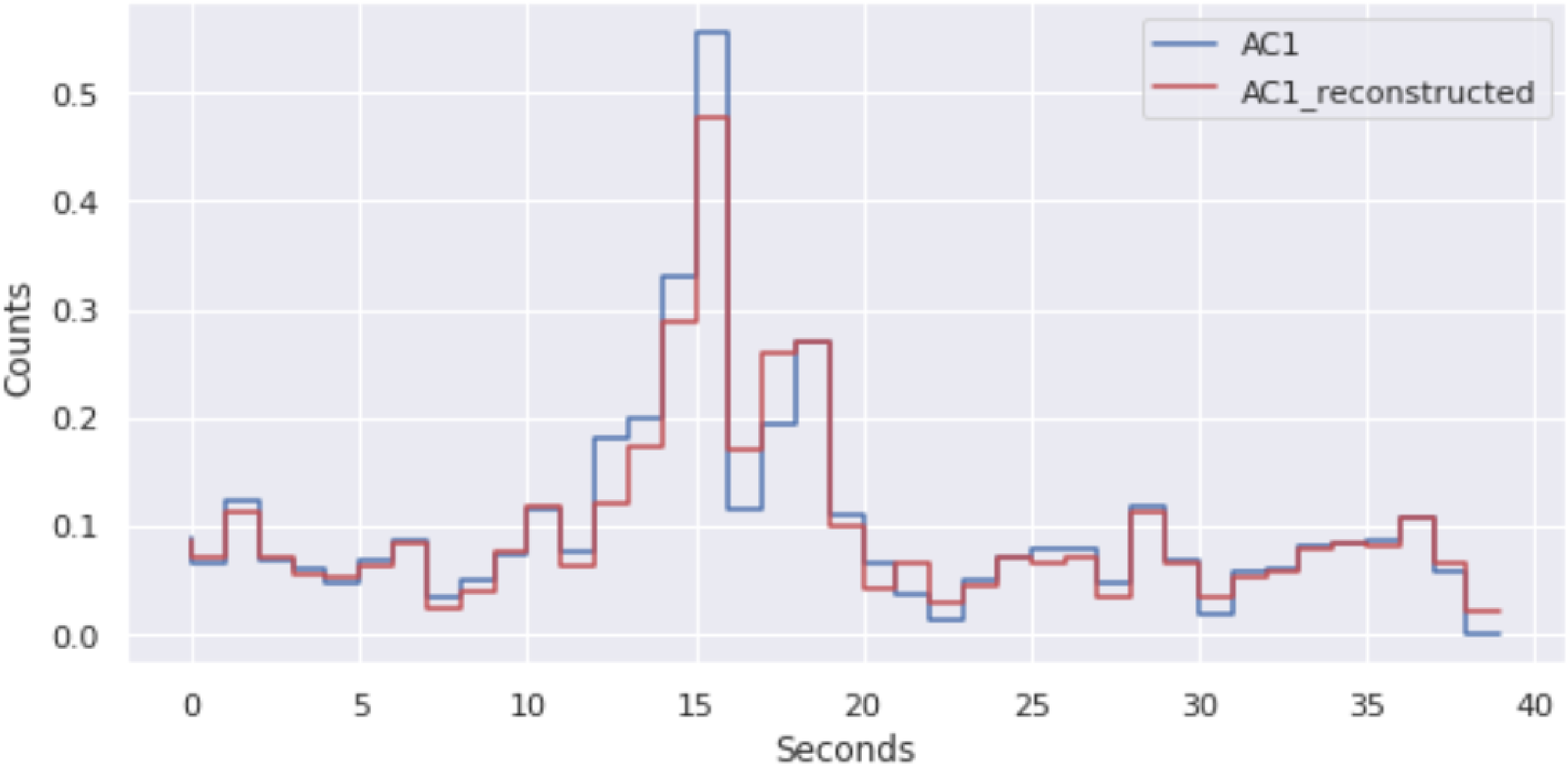}
	\caption{Example of time series of one of the ACS panels. The time series shown in the top image is part of the test dataset, while the bottom one is part of the MTSs containing GRBs used to evaluate the model. The blue line represents the original signal, while the red line indicates the reconstructed one. }
	\label{fig:architecture}
\end{figure*}

\section{Preliminary Results}

Once the threshold is defined, we evaluate the trained model by calculating the anomaly score of a list of MTSs containing GRBs (anomalies), previously identified in the AGILE ACS data by carrying out a cross-search with events detected by other space missions. In addition, to evaluate the false positive rate, we analyse the validation dataset composed of background-only MTS. The model correctly detected 82 GRBs over 85 from the GRB list with only one false positive over 749 MTSs in the validation dataset. Therefore, we can calculate the Precision and Recall values as: 

\begin{equation}
Precision = \frac{TP}{TP+FP} \text{ and } Recall=\frac{TP}{TP+FN}
\end{equation}
where TP = True Positive, FP = False Positive, and and FN = False Negative. We obtain a Recall of 0.96 and a Precision of 0.99, proving the capability of the model to detect GRBs in the ACS data.

\section{Conclusions and Future Works}

We developed a new DL model based on the CNN autoencoder architecture to detect GRBs inside the ACS data. The model is trained with a dataset of background-only MTSs randomly extracted from the ACS data. We evaluated the trained model using a list of MTSs manually selected containing confirmed GRBs, proving the capability of the model to detect GRBs inside the ACS data. This new model can be implemented into the AGILE real-time analysis pipeline to perform the follow-up of external science alerts, adding a new tool to analyse the data acquired by the ACS that was not originally designed to detect GRBs. 

In the future, we plan to optimise the hyperparameters (e.g. the number of layers, MTS time window size, etc), evaluate the false alarm rate of the model to obtain the statistical significance of detections, and use the model to search for GRBs in the ACS data using external GRB catalogues.

\acknowledgements The AGILE Mission is funded by the Italian Space Agency (ASI) with scientific and programmatic participation by the Italian National Institute for Astrophysics (INAF) and the Italian National Institute for Nuclear Physics (INFN). The investigation is supported by the ASI grant  I/028/12/6. We thank the ASI management for unfailing support during AGILE operations. We acknowledge the effort of ASI and industry personnel in operating the  ASI ground station in Malindi (Kenya), and the data processing done at the ASI/SSDC in Rome: the success of AGILE scientific operations depends on the effectiveness of the data flow from Kenya to SSDC and the data analysis and software management.


\bibliography{X1-007}


\end{document}